\documentclass[lettersize,journal]{IEEEtran}

\usepackage[ left=0.625in, top=0.75in, right=0.75in, bottom=1.2in, headsep=0.25in, headheight=20pt, heightrounded, letterpaper] {geometry}
\usepackage[utf8]{inputenc}
\usepackage{lipsum}
\usepackage{comment}
\usepackage{xfrac}  

\usepackage{lipsum,microtype}
\usepackage{authblk}
\usepackage[ruled,vlined]{algorithm2e}
\usepackage{amsmath, nccmath}
\usepackage{geometry}
\usepackage{tabularx}
\usepackage{booktabs}
\usepackage{setspace}
\usepackage{mathtools}
\usepackage{amssymb}
\usepackage{xcolor}
\usepackage{subcaption}
\usepackage{graphicx}
\graphicspath{{./Fig/}} 

\usepackage[ruled,vlined]{algorithm2e}
\usepackage{cite}
\newcounter{mycounter}
\usepackage{subfig} 
\usepackage{fancyhdr}
\usepackage{textcomp}
\usepackage{blindtext}
\usepackage{optidef}
\usepackage{algpseudocode}
\usepackage{amsmath}

\title{LLM-Enabled Data Transmission in End-to-End Semantic Communication}

\author{Shavbo Salehi\IEEEauthorrefmark{1}, Melike Erol-Kantarci\IEEEauthorrefmark{1}, \textit{Fellow, IEEE}, Dusit Niyato\IEEEauthorrefmark{2}, \textit{Fellow, IEEE}\\

\IEEEauthorblockA{\IEEEauthorrefmark{1}\textit{School of Electrical Engineering and Computer Science, University of Ottawa, Ottawa, Canada}}\\
\IEEEauthorblockA{\IEEEauthorrefmark{2}\textit{Nanyang Technological University, Singapore}}\\
\IEEEauthorblockA{Emails: \{ssale038, melike.erolkantarci\}@uottawa.ca, \\\ dniyato@ntu.edu.sg}}

\date{}
\begin{document}

\twocolumn

\maketitle
\begin{abstract}

Emerging services such as augmented reality (AR) and virtual reality (VR) have increased the volume of data transmitted in wireless communication systems, revealing the limitations of traditional Shannon theory. To address these limitations, semantic communication has been proposed as a solution that prioritizes the meaning of messages over the exact transmission of bits. This paper explores semantic communication for text data transmission in end-to-end (E2E) systems through a novel approach called KG-LLM semantic communication, which integrates knowledge graph (KG) extraction and large language model (LLM) coding. In this method, the transmitter first utilizes a KG to extract key entities and relationships from sentences. The extracted information is then encoded using an LLM to obtain the semantic meaning. On the receiver side, messages are decoded using another LLM, while a bidirectional encoder representations from transformers (i.e., BERT) model further refines the reconstructed sentences for improved semantic similarity. The KG-LLM semantic communication method reduces the transmitted text data volume by 30\% through KG-based compression and achieves 84\% semantic similarity between the original and received messages. This demonstrates the KG-LLM methods efficiency and robustness in semantic communication systems, outperforming the deep learning-based semantic communication model (DeepSC), which achieves only 63\%.
\end{abstract}

\begin{IEEEkeywords}
Knowledge Graph, LLM, LLM-Encoding, LLM-Decoding, Semantic Communication, SST2. \vspace{-0.3cm}
\end{IEEEkeywords}

\section{Introduction}

With the growing use of data-driven applications and the increasing demand for real-time services, the need to transition from traditional wireless paradigms to more advanced frameworks has become apparent. Traditional communication systems were designed to ensure accurate data transmission, focusing on data-level communication. However, as networks evolve toward 6G, the limitations of traditional methods, such as those based on Shannon theory, have become more evident, particularly in terms of physical capacity limits, low reliability, and high latency \cite{fernandes2024semantic}. Consequently, semantic communication proposed as a solution, especially in applications like extended reality (XR), where the meaning of transmitted information is more critical than its exact representation \cite{zhou2024goal}.

Semantic communication represents a shift in wireless communication, moving away from a traditional focus on raw data transmission toward an effective delivery of meaning and intent
. Semantic communication, utilizes advanced techniques such as natural language processing (NLP), machine learning (ML) and knowledge graphs (KGs) to extract and transfer the underlying semantics of messages \cite{liang2024generative}. By prioritizing meaning over exact data representation, semantic communication not only reduces data redundancy, but also improves resource allocation efficiency, adapts to varying contexts, and enables intelligent communication \cite{liang2024generative}. These features position semantic communication as a key enabler for next-generation wireless networks in applications that require real-time response, contextual awareness, and optimal utilization of constrained resources. 

Traditional ML models provide structured frameworks for understanding and modeling semantic communication, but they face limitations in scalability, adaptability, and contextual understanding \cite{getu2023tutorial}. Large language models (LLMs) address these gaps by processing 
data with deep contextual awareness \cite{yang2024rethinking}. The integration of LLMs into semantic communication enables advanced wireless systems to perceive intent, resolve ambiguities, and handle diverse scenarios. This leads to considering LLMs as a novel solution to realize the full potential of semantic communication
. However, existing semantic communications
lack theoretical guidance for semantic channel coding and struggle to handle inconsistent knowledge basis \cite{luo2022semantic}. 
It also has data heterogeneity and semantic ambiguity in multimodal systems \cite{jiang2024large}. To overcome these challenges, we propose KG-LLM semantic communication, a novel, three-step, LLM-based method for text data transmission in the semantic communication system. The proposed KG-LLM
framework enhances transmission efficiency by integrating KGs for structured compression and using LLMs for semantic encoding. This approach reduces the volume of transmitted text data while preserving essential semantic information, optimizing bandwidth usage. Additionally, the framework ensures high semantic fidelity through LLM-based encoding and BERT-driven refinement, making it 
capable of reconstructing meaningful messages with improved contextual understanding.

\begin{figure*}[htbp]
  \centering
  \includegraphics[width=0.9\textwidth, height=4.73cm]{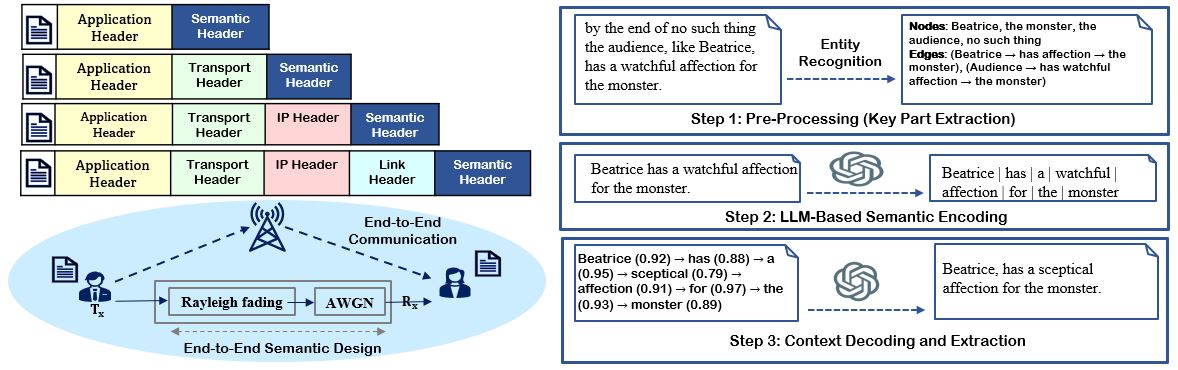} 
  \caption{System model of LLM-enabled semantic communication}
  \label{fig:sce1}
          \vspace{-0.5cm}
\end{figure*}

\section{Literature Review}

Traditional ML methods have been used in wireless semantic communication to integrate shared knowledge bases, thereby improving the effectiveness of semantic transmission. In particular, for small Internet of Things (IoT) devices with limited bandwidth and computational resources, reducing the transmitted text size while preserving semantic integrity enhances communication efficiency, making it more suitable for these scenarios. In \cite{wang2022performanceJSAC}, an attention-based learning approach was proposed to optimize semantic communication, highlighting the potential of deep learning (DL) to address the challenges of resource allocation and semantic fidelity. However, DL-based methods often lack semantic understanding, interpretability, and explainability \cite{getu2023tutorial}. The emergence of LLM has allowed wireless communication systems to overcome the shortcomings of DL-based methods, particularly in real-time scenarios \cite{guo2024large}. LLMs have shown potential in improving real-time wireless communications applications by enabling semantic understanding where it is most impactful \cite{jiang2024large}.


Empowering E2E semantic communication—which encodes, transmits, and decodes meaning directly—with LLMs overcomes DL-based limitations by preserving context, handling long-range dependencies, and adapting to diverse scenarios, whereas DL models rely on pattern recognition and lack semantic reasoning \cite{boateng2024survey}. Furthermore, these models have enabled advances in translating and compressing natural language information for wireless communications, mitigating the deficiencies of traditional ML models. In \cite{yang2024rethinking}, the challenges of multiuser access in 6G are addressed by a semantic communication framework for multimodal LLM, where shared knowledge bases enable standardized encoding and personalized decoding. Similarly, \cite{lu2024generative} proposes a framework to improve multimodal semantic communication, leading to optimal transmission, robustness, and resistance to noise in vehicular networks. In \cite{hello2024semantic}, an LLM-based method with graph neural networks (GNNs) is proposed for the semantic encoding of KGs in low-dimensional latent spaces and reconstructing the KG in the receiver using the GNN-based node and relation classification.

Despite their strengths, LLMs have limitations, primarily requiring computational resources, making them less practical for real-time applications in mobile and edge communications \cite{wang2024uses}. Moreover, their high resource consumption and security concerns 
pose challenges for deployment in resource-constrained wireless networks.
These weaknesses require compatible techniques to improve the functionality of LLMs in semantic communications \cite{chaccour2024less}. To overcome these shortcomings, we propose KG-LLM, a framework that integrates KG with LLMs. The GNN-based method in \cite{hello2024semantic} directly encodes the entire KG into a low-dimensional latent space, relying on GNNs for node and relation classification to reconstruct the KG at the receiver, while our KG-LLM model first extracts structured representations of semantic relationships, enhancing the contextual reasoning capabilities of the LLM. This approach reduces computational demands by dividing tasks between the KG and LLM. It also improves efficiency by using KGs for structured data processing while reserving LLMs for unstructured or complex contexts.

\section{System Model}

In this work, we explore an E2E semantic communication scenario in which two users, a transmitter (Tx) and a receiver (Rx), transmit messages through a gNodeB (gNB). Messages have two types, structured and unstructured; structured messages are summarized by a KG, with its key entities extracted and represented as triple expressions, while unstructured sentences remain unchanged. The output of this step is passed to the next processing layer, where the sentences are encoded using an LLM. The encoded messages are then transmitted over a wireless channel with the gNB which functions as a relay, forwarding the encoded messages to the Rx. On the Rx side, the encoded messages are decoded using a combination of LLM and BERT to reconstruct the message, focusing on increasing the semantic similarity between the original and the reconstructed messages. If the received semantic information is inaccurate or incomplete, a feedback mechanism can prompt the Tx to re-extract and retransmit the semantic content, ensuring reliability in communication. Using the BERT model, the system infers and restores missing parts of the message based on its context, ensuring high-fidelity reconstruction.

Fig. \ref{fig:sce1} illustrates the E2E scenario, where text data is transmitted over a Rayleigh fading channel with additive white Gaussian noise (AWGN) and 
multipath propagation. The figure also highlights the roles of the KG, the encoding and decoding processes of the LLM models, and the relay functionality of the gNB. Despite the advantages of semantic communication in transmitting the semantics of sentences, it introduces an overhead in the transmitted message \cite{gao2024cross}, adding approximately 10 bytes per packet due to the semantic processing layer compared to traditional packet transmission systems \cite{guo2024semantic}. 
This work focuses on the semantic aspects of communication, the integration of KG and LLMs for efficient text representation and reconstruction. Security-related aspects are not considered in this study, as they fall outside the intended scope.
        \vspace{-0.3cm}

\section{Proposed Method}

In this paper, we propose the KG-LLM semantic communication method for E2E semantic communication. The proposed method is evaluated on the Stanford Sentiment Treebank v2 (SST2) dataset, a sentiment-based dataset that includes both structured and unstructured sentences. The three-step semantic communication method, described in Algorithm 1, first separates structured and unstructured sentences. Then, KG extraction applies for structured sentences while skipping triple extraction for unstructured ones, as KG cannot process such sentences. After preprocessing, the output is encoded by an LLM and transmitted to the Rx, where the same LLM is used to decode the packet, ensuring synchronization between the Tx and Rx. The decoded message is then refined using a BERT model to enhance semantic accuracy.

\begin{algorithm}[t!]
\caption{KG-LLM semantic communication Framework}\label{alg:end_to_end}
\SetAlgoLined
\KwIn{Sentence \( S \)}
\KwOut{Reconstructed \( S_{\text{dec}} \)}
\vspace{5pt}

\textbf{Step 1: KG Semantic Extraction}\;
\Begin{
    Tokenize \( S \) into tokens \( T = \{t_1, t_2, \dots, t_n\} \)\;
    Calculate entropy (\(\mathbb{E}(S)\))\ for \( S \)\;
    Perform NER on \( S \) to extract entities \( \mathcal{E}_i = \{e_{i,1}, e_{i,2}, \dots, e_{i,j}\} \)\; 
\If{\(E(S) > H_{\theta} \)}{
    \(S \in \{S_j\}\)\;
}
\Else{
    \(S \in \{S_i\}\)\;
    Apply POS tagging\;
    Utilize Dependency Parsing\; 
    Generate a KG \( G_i = (V_i, E_i) \)\;
}
Aggregate \( \{S_j\} \cup \{S_i\} \) as \( S_{\text{tot}} \)\;
}
\textbf{Step 2: LLM-Based Semantic Encoding}\;
\Begin{
    Tokenize \( S_{\text{tot}} \) into tokens \( T_{\text{tot}} = \{t_1, t_2, \dots, t_i\} \) using BPE\;
    
    Encode \( T_{\text{tot}} \) using the LLM to obtain contextualized token representations:
    \[
    S_{\text{enc}} = \text{LLM}(T_{\text{tot}})
    \]
    Transmit \( S_{\text{enc}} \) over the wireless communication channel\;
}
\textbf{Step 3: Contextual Decoding and Refinement}\;
\Begin{
    Receive \( S_{\text{enc}} \)  at the Rx\;
    Decode \( S_{\text{enc}} \)  using the LLM:
    \[
    S_{\text{dec}} = \text{LLM}(S_{\text{enc}})
    \]
    Refine \( S_{\text{dec}} \) using BERT\;
}
        \vspace{-0.1cm}
\end{algorithm}

\subsection{Pre-processing with Knowledge Graph}
\label{subsection:step1}



In the first step, KG $G$ extracts triples from sentence $S$, including subjects, objects, and their relationships, to eliminate unnecessary information, bandwidth savings, and reduced system delays. For KG extraction, the method needs to identify key entities and their relationships. Tokenization is applied to divide each $S_i$ into words or phrases, which is \( T_i = \{t_{i,1}, t_{i,2}, \dots, t_{i,j}\} \), where $t_{i,j}$ denotes the token $j$-th of the sentence $S_i$. The named entity recognition (NER) detects entities in $S_i$, denoted as \( \mathcal{E}_i = \{e_{i,1}, e_{i,2}, \dots, e_{i,j}\} \), where $e_{i,j}$ is the detected entity $j$-th in $S_i$. Then a relationship extraction algorithm identifies connections between entities. In the extraction phase $\mathcal{E}_i$, nouns, noun phrases, and named entities are identified as nodes ${V_i}$, in $G$. The extraction of $G$ is followed by part-of-speech (PoS) tagging to identify grammatical roles. The dependency parsing is then used to analyze the syntactic relationships between entities. During the relationship extraction phase, verbs or action phrases are detected as relationships, while semantic role labeling or dependency parsing establishes connections between entities. To structure $G$, the resource description framework (RDF) and the Web ontology language (OWL) are used \cite{gangemi2023text2amr2fred}. Finally, in KG construction phase, entities as nodes ${V_i}$, and the relationships as directed edges ${E_i}$ are represented by $G_i = (V_i, E_i)$. 
 
The SST2 data set is designed as a sentiment data set in which there are structured ($S_i$) and unstructured sentences ($S_j$). $S_j$ are metaphors or unstructured sentences that lack an explicit structure. Since KG cannot extract meaningful triples for $S_j$, the first step is skipped to reduce computational complexity and avoid generating triples that do not accurately capture the meaning of $S_j$. In the KG-LLM method, $S_j$ are detected by entropy ($\mathbb{E}$) analysis on $S$ as follows:
\begin{equation}
     \mathbb{E}(S) = - \sum_{i=1}^N p(T_i) \log p(T_i)  
\end{equation}
where \( p(T_i) \) is the probability of token $T_i$ in \( S \), calculates the unpredictability of the sentence using word probability distributions, in which \( S \in S_j\) if \( \mathbb{E}(S) > H_{\text{$\theta$}} \), and $H_{\text{$\theta$}}$ is an entropy threshold. Finally, both $S_i$ and $S_j$ are aggregated into a unified set, defined as $S_{\text{tot}}=\{S_i,S_j\}$ and then sent to the next step for further processing.

\subsection{Semantic Encoding with LLM}

After KG triple extraction, the system encodes \( S_{\text{tot}}\) using the LLM for transmission. Using the byte pair encoding (BPE) tokenizer, the input \( S_{\text{tot}} \) is tokenized into smaller units $T_{\text{tot}} = \{t_1, t_2, \dots, t_i\}$, forming the following token set:

\begin{equation}
    \forall t_i \in {T_{\text{tot}}},     em_i = f_{emb}(t_i)\in \mathbb{R}^d
\end{equation}
where $T_{\text{tot}}$ represents the full sequence of tokens generated from \( S_{\text{tot}} \) after tokenization. This process ensures that each token $t_i \in {T_{\text{tot}}}$ is assigned a learned embedding $em_i \in \mathbb{R}^d$ through the model embedding layer, which captures the basic semantic and syntactic features of each $t_i$. Then the embeddings $em = \{em_1, em_2, \dots, em_i\}$ are processed by the LLM encoder transformer layers. The LLM employs a sequence-to-sequence (seq2seq) architecture, in which the encoder captures contextual dependencies between $T_{\text{tot}}$ and produces an intermediate representation. This representation is then passed to the decoder on the Rx side, which generates an optimized semantic representation of $S$. Note that the LLM maintains token-level contextual representations throughout the encoding process. This ensures that long-range dependencies and relationships between tokens remain in the final representation.

After contextualization, the LLM output is a compressed sequence of tokens, represented as $S_{\text{enc}} = \{t'_1, t'_2, \dots, t'_m\}, \quad m \leq n$, where the length \( m \) of the encoded sequence ($S_{\text{enc}}$) is generally shorter than the original sentence length \( n \), leading to compression. Using LLMs encoding leads to decreasing the redundancy in messages for words that do not add semantic value. Furthermore, LLM-based encoding models compress the messages and transmit a sequence that preserves meaning that is a compact version of the original $S$. It should be noted that the effectiveness of compression depends on the complexity of $S$ since sentences with more entities and relationships generally can be less compressed. 

Finally, the encoded sequence $S_{\text{enc}}$ is transmitted over the wireless channel. The LLM-based encoding process ensures that the transmitted representation retains the most relevant semantic information, enabling efficient transmission in bandwidth-constrained environments.

\subsection{Contextual Inference with LLM and BERT Embeddings}


After receiving the encoded data on the Rx side, the system employs an LLM decoder to generate token probabilities that correspond to the most likely sequence of words given the transmitted representation. This step involves sequence-to-sequence mapping, where the LLM uses its pre-trained knowledge to reconstruct the original semantic structure while mitigating transmission-induced distortions. Additionally, attention mechanisms within the LLM allow it to dynamically focus on relevant contextual cues, ensuring that key semantic elements are preserved. To further refine the reconstructed text, a BERT-based embedding model is applied to correct potential errors and fill in missing contextual details. The objective is to generate a deeper understanding of the text by capturing meanings, relationships, and insights.

In this step of the KG-LLM semantic communication method, $S_{\text{enc}}$ is passed through a decoding layer in the LLM to map them back to token embeddings, $E_{\text{dec}} = \{e_1', e_2', \dots, e_n'\}$. These decoded embeddings are used to reconstruct the sentence \( S_{\text{dec}} \) and then converted into words, and similar words or phrases are placed closer together. The model requires semantic understanding, where embeddings capture relationships between words, enabling it to recognize concepts such as synonymy and polysemy. After that, the inference mechanism is implemented using embeddings to infer various aspects such as topic relevance, sentiment, intent, and even unspoken assumptions. Then BERT captures context in a bidirectional way by analyzing the entire sentence at once, looking at both the words before and after each word. This bidirectional approach helps them understand the relationships between words and phrases more deeply. It should be noted that the overall complexity of the KG-LLM model is $O(N^2d)$, where $d$ is the hidden dimension of the LLM.

\section{Simulation Results}

In this paper, the performance of the proposed approach is evaluated under different channel conditions by varying the signal-to-noise ratio (SNR), ranging from 2dB to 10dB. To find the most suitable LLM for KG-LLM semantic communication for encoding and decoding, we evaluated the performance of three versions of the T5 model (text-to-text transfer transformer) T5-small, T5-base, and T5-large on the SST2 dataset. T5-small has approximately 76 million parameters, T5-base has 247 million, and T5-large has 783 million parameters. To evaluate KG-LLM semantic communication, we compare its performance with the DL-based semantic communication model (DeepSC) \cite{xie2020lite} and the generative pre-trained transformer 2 (GPT2) \cite{wang2024uses}. In the KG-LLM method, the threshold is determined as $H_{\theta}=3.85$ based on our analysis and evaluation. All methods are implemented in Google Colab by Python 3.11.

\subsection{Semantic Compression Rate}

In the KG-LLM semantic communication method, due to the summarization of $S$ with KG, the size of the text messages in the SST2 data set is reduced by approximately 30\% for $S_i$. It should be noted that the KG-LLM semantic communication method skipped the KG extractions for $S_j$. The proposed method benefits from more expression because of the use of LLMs for encoding. LLMs focus on extracting the semantic meaning of a sentence rather than representing the exact text, $S_{tot}$. This step reduces redundancy by concentrating on key information, resulting in a compressed representation, which is called the compression ratio. In this paper, we calculate the compression ratio $CR$, using the following equation:

\begin{equation}
    CR = 1-\frac{S_{enc}}{S_{tot}}
\end{equation}
where for $CR<1$, the compressed message is shorter than the original message, indicating effective compression, $CR=1$ indicates no compression achieved, and $CR>1$ shows that the output message is longer than the original. This situation, $CR>1$, happened due to that the LLM rephrases or generates verbose output while attempting to maintain semantic accuracy due to the noise of the system. As shown in Fig. \ref{fig:FLANs}, Flan-T5-base $CR = 0.37$, which has significant compression, reducing the sentence length to ~37\% of its original length. Then T5-small, with $CR = 0.89$ has the less compression ratio, by saving about 90\% of the message. In the least $CR$ level T5-large is located with $CR=1.3151$, which expanded the message to ~131\% by adding redundant or explanatory phrases.
        \vspace{-0.4cm}

\subsection{Semantic Similarity}

\begin{figure*}
\centering
\begin{subfigure}{0.37\textwidth}
    \includegraphics[width=\textwidth]{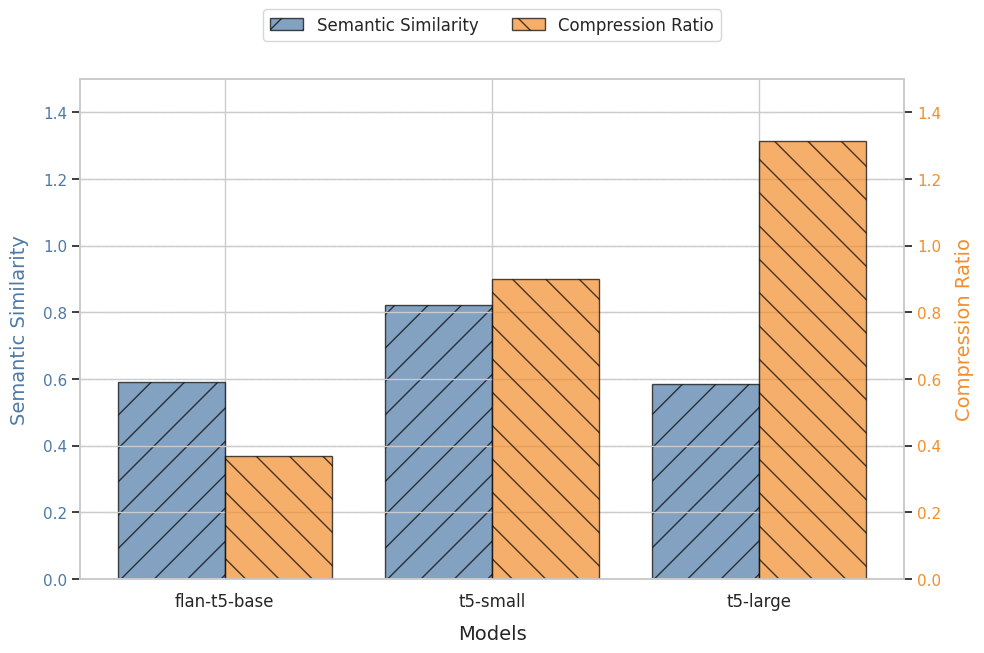}
    \caption{Flan-T5-base, T5-small, and T5-large performance as the LLMs for Encoding Decoding in semantic communication}
    \label{fig:FLANs}
\end{subfigure}
\hfill
\begin{subfigure}{0.3\textwidth}
    \includegraphics[width=\textwidth]{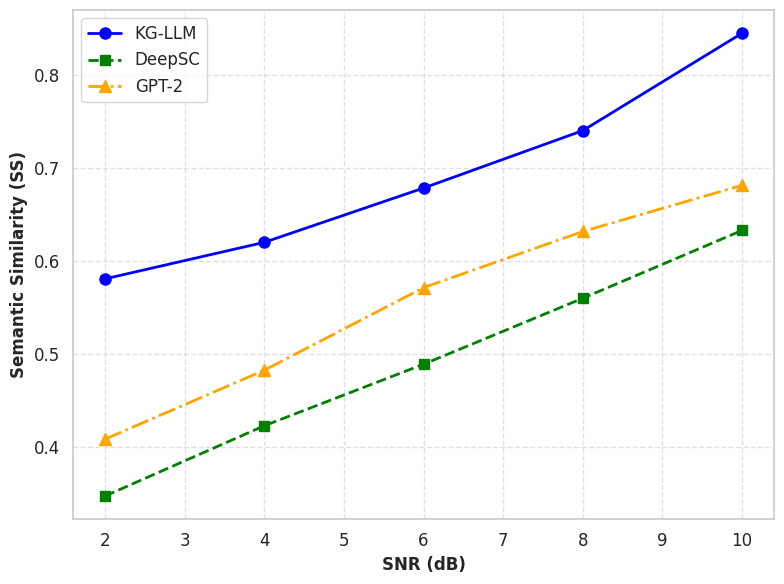}
    \caption{SS vs SNR for KG-LLM, DeepSC, and GPT-2}
    \label{fig:SS2}
\end{subfigure}
\hfill
\begin{subfigure}{0.3\textwidth}
    \includegraphics[width=\textwidth]{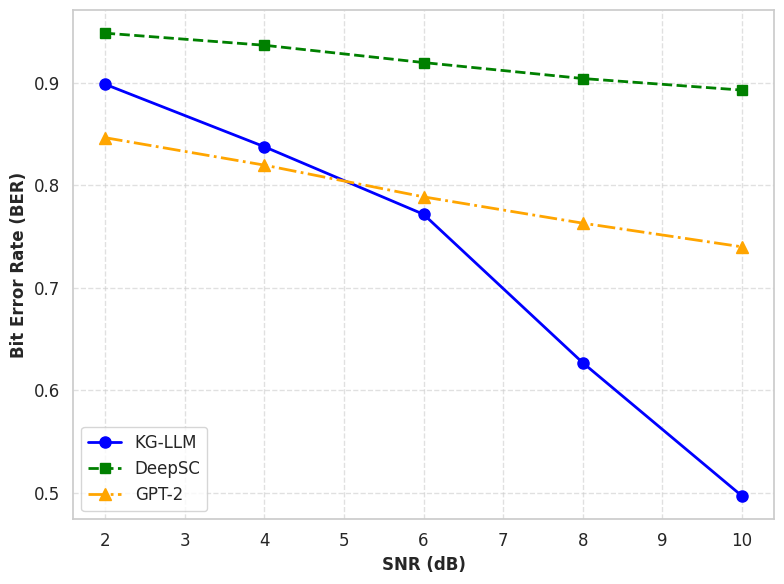}
    \caption{BER vs SNR for KG-LLM, DeepSC, and GPT-2}
    \label{fig:BER}
\end{subfigure}
\caption{Comparison of LLM-Based Semantic Communication Performance}
\label{all}
\vspace{-15pt}
        \vspace{-0.2cm}
\end{figure*}

In semantic communication, since $S$ is not transmitted, the models should concentrate on semantic similarity, $SS$, between $S$ in Tx and $S_{dec}$ on the Rx side. For this reason, $SS$ is one of the most important criteria in the semantic communication which is calculated by the following equation:

\begin{equation}
    SS = cos(V_{S},V_{S_{dec}})
\end{equation}
where $V_{S}$ and $V_{S_{dec}}$ are corresponding embedding vectors in a high-dimensional space. As shown in Fig. \ref{fig:FLANs}, the Flan-T5 base with $SS=0.59$, has a moderate similarity, indicating that while the core meaning of the message is preserved, due to compression, some parts of the meaning are lost. T5-small with $SS=0.83$ has the highest similarity, showing preserving the meaning alongside performing a slight compression. T5-large with $SS = 0.58$ has the same performance as Flan-T5-base, in which, while the expanded sentence retained some meaning, it deviated from the original in structure and semantic meaning.

\subsection{Transmission Time}

One of the objectives of semantic communication is to overcome the latency issues of traditional communications. While reducing the size of transmitted data is a key factor, transmission time ($TT$) is influenced not only by data compression but also by processing delays from semantic encoding, decoding, and potential retransmissions due to semantic errors. To assess overall efficiency, we evaluated ($TT$) as a performance criterion under various LLMs, considering both compression effects and computational overhead.
Based on our simulation results, the Flan-T5 base $TT=2.2249 ms$ shows that LLM has moderate processing and transmission time, with balancing complexity and efficiency. T5-small with $TT=1.6403 ms$ has a low processing time due to the simplicity and smaller size of the model. However, T5-large with $TT=20.7711 ms$ has the highest processing time, compared to the other algorithms considered, due to increased computational requirements.


Among the LLMs considered, the Flan-T5 base achieves the best compression at the cost of some $SS$. T5-small strikes a balance between compression and $SS$, making it efficient for practical scenarios. T5-large sacrifices compression for verbosity, leading to higher transmission costs and marginal semantic gains. The selection of a suitable LLM for semantic communication depends on the application. High compression scenarios prioritize compactness, and it is recommended to use Flan-T5-base, while high $SS$ scenarios prioritize accuracy and use T5-small, avoid models that expand sentences unnecessarily, and T5-large could be utilized. In KG-LLM, T5-small is selected for encoding and decoding. As we mentioned above, the overall complexity of the KG-LLM model is $O(N^2d)$, where $d=512$ for the T5-small, which dominates the encoding and decoding processes.
\vspace{-0.4cm}

\subsection{BER and Semantic Similarity Across SNR Levels}

In this section, to evaluate the KG-LLM method, we compare its performance with DeepSC and GPT-2. As depicted in Fig. \ref{fig:SS2}, $SS$ of KG-LLM increases 
by enhancing the SNR of the system, like DeepSC and GPT2. KG-LLM outperforms both DeepSC and GPT-2 at all levels of SNR, with its $SS$ increasing from 0.5812 at 2 dB to 0.8455 at 10 dB. The variation in $SS$ at different SNR levels is due to the impact of noise on semantic reconstruction—at lower SNRs, higher error rates degrade semantic integrity, while at higher SNRs, reduced noise allows for more accurate semantic preservation and improved decoding quality. This indicates that KG-LLM captures semantic meaning more effectively, even under low-SNR conditions. Contextual inference with LLM and BERT embeddings enables systems to gain an understanding of text. In contrast, DeepSC has the lowest $SS$, ranging from 0.3476 at 2 dB to 0.6334 at 10 dB, highlighting its limited ability to preserve semantic meaning under noise. GPT-2 shows better performance than DeepSC but falls short of KG-LLM, with semantic similarity ranging from 0.4088 to 0.6819.

Fig. \ref{fig:BER} shows the bit error rate (BER) for the three considered methods. KG-LLM has the highest performance, with a considerable reduction in BER as the SNR improves, dropping from 0.8988 at 2 dB to 0.4967 at 10 dB. GPT-2 performs better than DeepSC, starting at 0.8466 at 2 dB and reducing to 0.7400 at 10 dB. 
It should be noted that the lower BER performance of KG-LLM in the low SNR regime is due to its reliance on KG-based compression, which reduces redundancy in the transmitted representation. While compression is not affected by channel quality, the lack of redundancy makes encoded data more vulnerable to noise, leading to higher decoding errors at low SNR levels.
Meanwhile, DeepSC consistently exhibits the highest BER values, beginning at 0.9486 and only marginally decreasing to 0.8930 at 10 dB, indicating its weaker resistance to noise in the wireless communication channel. KG-LLM semantic communication achieves the best performance in the given scenario, with a BER of 0.9 at 2 dB SNR, significantly improving to 0.5 at 10 dB SNR.

Contextual inference with LLM and BERT embeddings has several advantages. These models go beyond surface-level text processing to provide a deeper understanding of context, which is essential for complex NLP tasks in 6G wireless communication networks. Furthermore, both embeddings are scalable, enabling real-time processing of large text volumes, making them ideal for high-throughput applications. 

\section{Conclusion}

In conclusion, the KG-LLM semantic communication framework demonstrates advances in efficiency, resilience, and semantic similarity for E2E communication systems. By KG-based preprocessing, the system reduces redundant information, leading to optimization in bandwidth utilization. Then the integration of LLMs enhances the system's ability to generate contextually relevant and concise outputs, making it highly suitable for transmission in noisy environments. The results highlight the superior performance of KG-LLM semantic communication compared to DeepSC and GPT-2, achieving higher semantic similarity and lower BER at higher SNR levels.
KG-LLM framework by using contextual inference through the combination of LLM and BERT embeddings, can accurately understand the meaning of text, making it a powerful tool for other applications that require deep contextual understanding.

\section*{Acknowledgment}
This work has been supported by the NSERC Canada Research Chairs program and Collaborative Research and Training Experience (CREATE) TRAVERSAL program under Grant 497981.

\vspace{-0.2cm}

\bibliographystyle{unsrt}
\bibliography{references}

\end{document}